# Heteroepitaxial growth of tetragonal $Mn_{2.7-x}Fe_xGa_{1.3}$ (0 ≤ x ≤ 1.2) Heusler films with perpendicular magnetic anisotropy

Adel Kalache[1], Anastasios Markou[1*], Susanne Selle[2], Thomas Höche[2], Gerhard H. Fecher[1], Claudia Felser[1*]

[1] Max Planck Institute for Chemical Physics of Solids, Nöthnitzer Str. 40, 01187, Dresden, Germany

[2] Fraunhofer Institute for Microstructure of Materials and Systems IMWS, 06120, Halle, Germany

**Abstract**

This work reports on the structural and magnetic properties of $Mn_{2.7-x}Fe_xGa_{1.3}$ Heusler films with different Fe content x (0 ≤ x ≤ 1.2). The films were deposited heteroepitaxially on MgO single crystal substrates, by magnetron sputtering. $Mn_{2.7-x}Fe_xGa_{1.3}$ films with the thickness of 35 nm were crystallized in tetragonal $D0_{22}$ structure with (001) preferred orientation. Tunable magnetic properties were achieved by changing the Fe content x. $Mn_{2.7-x}Fe_xGa_{1.3}$ thins films exhibit high uniaxial anisotropy $K_u \geq 1.4$ MJ/m$^3$, coercivity from 0.95 to 0.3 T and saturation magnetization from 290 to 570 kA/m. The film with $Mn_{1.6}Fe_{1.1}Ga_{1.3}$ composition shows high $K_u$ of 1.47 MJ/m$^3$ and energy product $(BH)_{max}$ of 37 kJ/m$^3$, at room temperature. These findings demonstrate that $Mn_{2.7-x}Fe_xGa_{1.3}$ films have promising properties for mid-range permanent magnet and spintronic applications.

## 1. Introduction

Currently, intensive efforts are made to develop novel permanent magnets in order to reduce or even completely replace their rare earth content (Nd, Sm).[1-3] The increasing global demand for permanent magnets has been driven by the development of high-efficiency motors and generators for various clean energy applications, such as wind turbines for power generation, electrical vehicles, and magnetic refrigeration. Therefore, the replacement of rare earth in permanent magnets became necessary, due to the volatility of their prices and the strategic issues associated with them. Moreover, new permanent magnets need to contain environmentally sustainable elements with lower environmental impact than rare earths. Apart from high-end applications, new magnets with mid-range performance are required as well, where novel hard magnetic materials are predicted to bridge the gap between low-cost hard ferrites and expensive rare-earth-based magnets.[2] To achieve this aim, the new class of permanent magnets need to meet the criteria of high Curie temperature ($T_c$) and high saturation magnetization ($M_s$) combined with strong uniaxial magnetic anisotropy ($K_u$), in order to achieve a large energy product value *(BH)$_{max}$* at room temperature (RT).[4]

---

[*] Authors to whom correspondence should be addressed: Anastasios.Markou@cpfs.mpg.de and Claudia.Felser@cpfs.mpg.de

Different approaches have been proposed to develop rare earth free permanent magnets. One of them is to induce magnetocrystalline anisotropy through tetragonal distortion in phases possessing high magnetization, such as $Fe_xCo_{1-x}$ alloy films grown on an appropriate buffer layer,[5-7] or doped by a third element[8-11]. Another promising approach is the exchange-coupled nanocomposites, where one phase with large coercivity $H_c$ is combined with a high-magnetization phase $M_s$.[12-15] The third involves searching and investigating new magnetic compounds exhibiting high magnetocrystalline anisotropy. As an example, it has been reported that the novel non-cubic $Zr_2Co_{11}$ and $HfCo_7$ alloys show attractive hard magnetic properties with large anisotropy.[16,17] Among new materials, Mn-based magnetic compounds[18,19] have attracted much interest in recent years, due to their high magnetocrystalline anisotropy, such as the tetragonal $Mn_xGa$,[20,21] $MnAl$[22] and the hexagonal $MnBi$.[23,24]

Heusler compounds are a remarkable class of materials with tunable multifunctional properties and a huge potential for various applications.[25,26] Tetragonally distorted D0$_{22}$ Mn-based Heusler compounds, such as the ferrimagnetic $Mn_xGa$ and $Mn_xGe$ (x = 2–3) show large magnetocrystalline anisotropy and large coercivity, as well as high Curie temperature.[19,27-30] However, they suffer from low saturation magnetization, due to the antiparallel coupling of the magnetic moments in the Mn atoms at 4d (Mn-II) and 2b (Mn-I) sites.[27,31] Therefore, the magnetization in the Mn-based tetragonal Heusler compounds has to be enhanced for the potential permanent magnets applications. Towards to this direction, it was reported that the tetragonal phase of the $Mn_{38.5}Fe_{32.5}Ga_{29}$ compound shows a coercivity of 0.33 T and a remanent magnetization of 20 Am$^2$kg$^{-1}$ at RT.[32] $Mn_{3-x}Y_xGa$ thin films with perpendicular magnetic anisotropy and tunable magnetization may be realized by the substitution of Mn by ferromagnetic elements (Y=Co, Fe).[33-35] This tunable magnetic behavior make these materials interesting for spintronic applications as well.

In this work, we present the structural and magnetic properties of tetragonally distorted $Mn_{2.7-x}Fe_xGa_{1.3}$ films with strong perpendicular magnetic anisotropy. For this purpose, we performed systematic X-ray diffraction (XRD), transmission electron microscopy (TEM) and magnetic characterization of films, heteroepitaxially grown on MgO substrates.

## 2. Experimental details

$Mn_{2.7-x}Fe_x$-$Ga_{1.3}$ (x = 0–1.2) films with thickness of 35 nm have been deposited on single crystal MgO (100) substrates. For the deposition, a BESTEC UHV magnetron sputtering system has been used with Mn (2"), Fe (2"), and $Mn_{50}Ga_{50}$ (2") sources in confocal geometry. The target to substrate distance was 17 cm. Prior to the deposition, the chamber was evacuated to a base pressure less than $2 \times 10^{-8}$ mbar, while the process gas (Ar 5N) pressure was $3 \times 10^{-3}$ mbar. Mn-Fe-Ga films were grown by co-sputtering at 350 °C, and then post-annealed in situ for additional 20 minutes to improve chemical ordering. The power of the different sources was adjusted, in order to reach the desired composition of the Mn-Fe-Ga films. Special attention was paid to keep Ga content equal for consistency, and to investigate the substitution of Mn by Fe. All samples were capped at room temperature with a 2 nm thick Al film to prevent oxidation. Stoichiometry was estimated by energy-dispersive X-ray analysis (EDX) and verified by inductively coupled plasma optical emission spectrometry (ICP-OES). X-ray diffraction (XRD) was collected with a Panalytical Diffractometer X'PERT$^3$ MRD, by using Cu-K$\alpha_1$ radiation ($\lambda$=1.5406 Å). The film thicknesses were determined by X-ray reflectivity (XRR) measurements. Transmission electron microscopy (TEM) and scanning transmission electron microscopy (STEM) were performed by a FEI Tecnai G2 F20 microscope at 200 kV and a FEI TITAN$^{-3}$ G2 80-300 microscope equipped with a SuperX energy-dispersive X-ray spectroscopy (EDXS) analyzer at 300 keV, respectively. TEM samples were prepared by focused ion beam milling (FIB). A protective Pt layer was deposited before the cross sectioning. Magnetic measurements were carried by out using a Quantum Design (MPMS 3) magnetometer.

## 3. Results and discussion

Different XRD measurements, such as θ–2θ, rocking curve, and phi-scans were performed to study the structure, the crystallinity, and the heteroepitaxial relationship between the films and the substrate. XRD patterns of the films with different Fe content x are shown in Fig. 1 (a). All samples have been indexed by assuming a tetragonal $D0_{22}$ phase. In the tetragonal D0$_{22}$ cell of stoichiometric Mn$_3$Ga, Mn, and Ga occupy two and one crystallographic sites, respectively. The resulting space group is *I4/mmm* with MnII on 4*d* (0, 1/2, 1/4), MnI on 2*b* (0, 0, 1/2), and Ga on 2*a* (0, 0, 0), as depicted in Fig. 2 (b). Fe in Mn$_{2.7-x}$Fe$_x$Ga$_{1.3}$ is expected to be preferentially substituted on the 4d site, due to its higher electronegativity, as previously known for Heusler compounds.[36] However, the neighboring atomic number of Fe and Mn most probably induce some chemical disorder, inducing a statistical replacement on both Mn positions. Only the (002) and (004) reflections were observed in XRD patterns, which indicates that the five investigated samples are crystallized in the tetragonal D0$_{22}$ structure with (001) preferred orientation and the c-axis normal to the film plane. Furthermore, the reflections of the Mn$_{1.5}$Fe$_{1.2}$Ga$_{1.3}$ film display some broadening. This can be attributed to a chemical disorder, which degrades the crystalline quality of the film, compared to films with intermediate composition. Moreover, the shoulder close to the (004) reflection can be attributed to a Mn-Fe chemical disorder, rather than to a pseudocubic[36] or coexisting cubic phase,[35] which have been observed in similar systems. Phi scan patterns of the {112} planes for the Mn$_{1.7}$Fe$_1$Ga$_{1.3}$ film and the {111} planes for the MgO substrate are depicted in Fig. 1 (c). The four peaks of Mn$_{1.7}$Fe$_1$Ga$_{1.3}$ show four fold symmetry with 90° intervals, suggesting a single crystalline film, with well-defined in-plane orientation. The two different sets of lattice planes occur at the same azimuthal angle φ, which indicates that the unit cell of the Mn$_{1.7}$Fe$_1$Ga$_{1.3}$ film is well aligned in the basal planes of the MgO substrate with cube-on-cube growth. Therefore, the heteroepitaxial relationship is Mn$_{1.7}$Fe$_1$Ga$_{1.3}$[100](001)∥MgO[100](001).

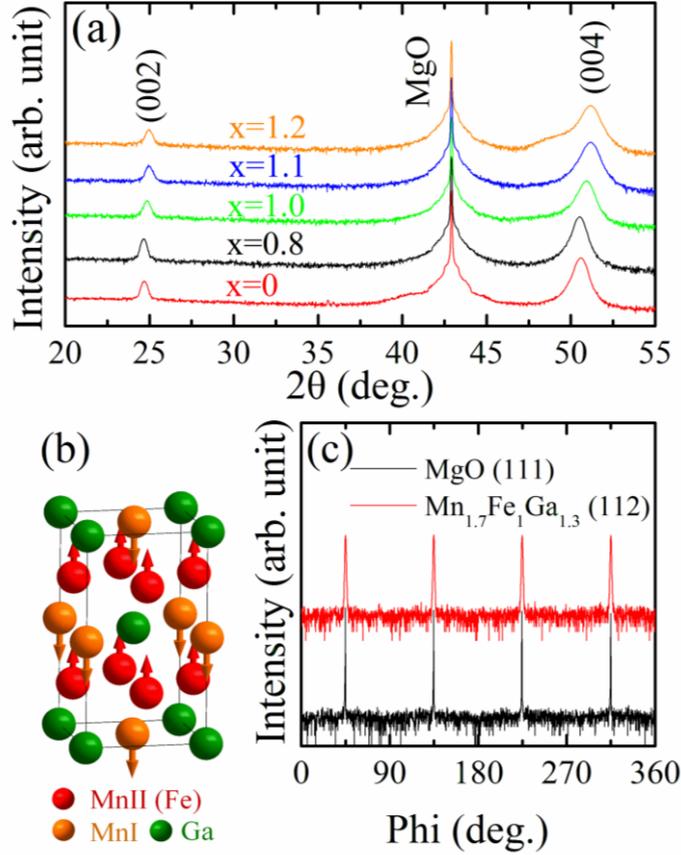

FIG. 1 (a) XRD patterns of $Mn_{2.7-x}Fe_xGa_{1.3}$ films with different Fe content x. (b) The tetragonal $D0_{22}$ unit cell of $Mn_3Ga$ with MnII (red) on 4d, MnI (orange) on 2b, and Ga (green) on 2a and the expected preferential occupancy of the substituted Fe on 4d. Arrows on Mn atoms indicate the antiparallel coupling of their magnetic moments, leading to ferrimagnetic ordering. (c) Phi scan measurements of the {111} and {112} planes from the MgO substrate and $Mn_{1.7}Fe_1Ga_{1.3}$ film, respectively.

To understand the effect of the Fe on the structure of $Mn_{2.7-x}Fe_xGa_{1.3}$ films, the lattice parameters, the *c/a* ratio, and the full width at half maximum (*FWHM*) of the (004) out-of-plane rocking curve are summarized in Table I. Lattice parameter *c* is deduced from the (002) and (004) reflections from XRD patterns, while lattice parameter *a* is calculated from the (112) reflection, as suggested in Ref. 37. Starting from the sample without Fe (x = 0) to x = 0.8, lattice parameters *c* and *a* remain almost constant. From x = 0.8 to x = 1.2 both lattice parameters decrease, while the *c/a* ratio slightly increases, which implies that the tetragonal distortion is not influenced significantly by the Fe. The broadening of the rocking curves with increasing Fe, is mainly caused by the Mn-Fe chemical disorder or/and the presence of defects.

TABLE I. Lattice parameters $c$ and $a$, $c/a$ ratio, and the full width at half maximum (*FWHM*) of the (004) rocking curve profile from $Mn_{2.7-x}Fe_xGa_{1.3}$ films.

| $Mn_{2.7-x}Fe_xGa_{1.3}$ | $c$ (Å) | $a$ (Å) | $c/a$ ratio | *FWHM* (deg.) |
|---|---|---|---|---|
| x=0 | 7.206 | 3.931 | 1.833 | 0.971 |
| x=0.8 | 7.209 | 3.930 | 1.834 | 1.063 |
| x=1 | 7.161 | 3.903 | 1.836 | 1.103 |
| x=1.1 | 7.137 | 3.883 | 1.838 | 1.152 |
| x=1.2 | 7.136 | 3.881 | 1.839 | 1.221 |

The heteroepitaxial relationship between the film and substrate was further confirmed by the TEM analysis. Fig. 2 (a) shows the cross-sectional STEM image of the $Mn_{1.5}Fe_{1.2}Ga_{1.3}$ film, where the film, the MgO substrate, and the protective Pt layer are clearly shown in different brightness. The $Mn_{1.5}Fe_{1.2}Ga_{1.3}$ film is continuous and relatively smooth with a thickness of 35 nm. The HRTEM image, close to the interface of the $Mn_{1.5}Fe_{1.2}Ga_{1.3}$ film and the MgO substrate is depicted in Fig. 2 (b). The matching of atomic planes across the interface between the $Mn_{1.5}Fe_{1.2}Ga_{1.3}$ film and the MgO substrate verifies the heteroepitaxial growth. The selected area electron diffraction pattern (SAED) of the same sample is depicted in Fig. 2 (c), where the electron beam is parallel to the [100] zone axis of the film and the substrate. Some low intensity small diffraction spot arise from the crystalline Pt layer. The two different sets of diffraction spots are aligned, confirming the cube-on-cube heteroepitaxial relationship between the $Mn_{1.5}Fe_{1.2}Ga_{1.3}$ and the MgO substrate. The indexed SAED pattern (Fig. 2 (d)) reveals that the film is crystallized in the tetragonal $D0_{22}$ crystal structure. Furthermore, the SAED pattern confirms that there is no other phase present in the $Mn_{1.5}Fe_{1.2}Ga_{1.3}$ film and the shoulder appearing close to the (004) reflection in XRD pattern (Fig.1 (a)) is due to Mn-Fe chemical disorder. The lattice constants were found to be $c = 7.139$ Å and $a = 3.889$ Å, which are in a good agreement with the XRD measurements. This small difference can be attributed to the higher measurement uncertainty, when deducing the lattice constants from the electron diffraction pattern.[38]

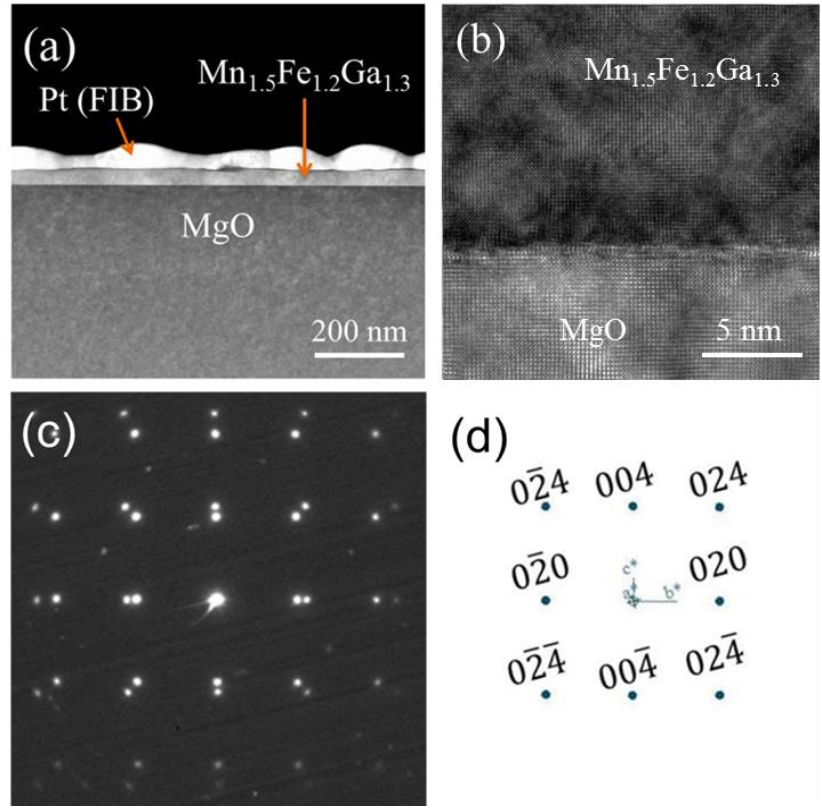

FIG. 2 (a) Cross sectional STEM image, (b) HRTEM image, (c) selected area diffraction pattern, and (d) simulated pattern of the $Mn_{1.5}Fe_{1.2}Ga_{1.3}$ film. The arrows in (a) indicate the location of the Heusler film and the Pt layer, which protected the film from the FIB erosion during the TEM sample preparation.

The $Mn_{1.5}Fe_{1.2}Ga_{1.3}$ film was investigated by cross sectional HAADF-STEM, and elemental mapping was realized using EDXS, as depicted in Fig. 3. The spatial distributions of the count rate intensity of the Mn, Fe, and Ga elements are represented independently with different colors in Fig. 3 (a) - (c). The three elements are homogeneously distributed over the entire film, confirming that the $Mn_{1.5}Fe_{1.2}Ga_{1.3}$ film is uniform and single phase. For further analysis, line scans along the film and across the film are depicted in Fig. 3 (e) and (f), respectively. The quantification of Mn, Fe, and Ga atomic ratios presents homogeneous distribution, apart from minor deviations at the interfaces in the line scan across the film. Therefore, it can be concluded that Mn was successfully substituted by Fe, without segregation or formation of a minority phase. The quantified line scans show a slightly higher Fe content in comparison to Ga, which can be attributed to the incomplete deconvolution of the Fe-$K_\alpha$ and Mn-$K_\beta$ edges, leading to an overestimation of the Fe content, as compared to ICP-OES results. At the MgO substrate-film interface, the Mn content seems to rise earlier, suggesting a favored accumulation of Mn atoms in the first monolayers of the film. This Mn enriched area is less than 1 nm, and might be attributed to the Mn affinity with O originating from MgO. This feature can be improved by the use of an appropriate buffer layer. The same applies to the film-Al interface capping, where the Mn content decreases slower than that of the Fe and Ga, as the Pt content increases. Therefore, this small intermixing with Pt can be attributed to the procedure of the FIB preparation. Additionally, the line scan across the interface shows O content of 5 to 10 at.% within the film, due to the oxidation after the sample preparation, as the cross section is necessarily exposed to the air, during the transfer from the FIB device to the microscope. Nevertheless, Mn, Fe and Ga elements are uniformly distributed over nearly the entire $Mn_{1.5}Fe_{1.2}Ga_{1.3}$ film.

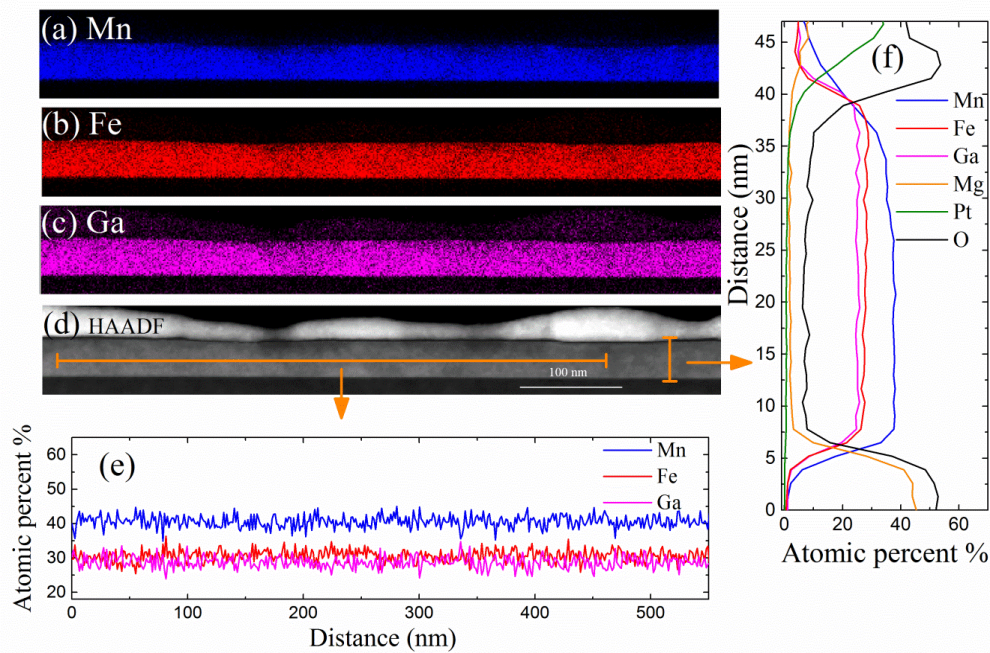

FIG. 3 Elemental mapping of (a) Mn, (b) Fe, (c) Ga and the corresponding (d) HAADF image of the $Mn_{1.5}Fe_{1.2}Ga_{1.3}$ film. Quantification of the Mn, Fe, and Ga content (e) along the film and (f) across the film. The orange bars in the HAADF image indicate the position of the line scans.

Typical in-plane and out-of-plane magnetization hysteresis loops of the $Mn_{1.9}Fe_{0.8}Ga_{1.3}$ film, measured at 300 K are shown in Fig. 4 (a). The perpendicular magnetic anisotropy is manifested by the easy saturation axis oriented along the film normal. The out-of-plane coercive field is 0.78 T and the saturation magnetization is 390 kA/m. The shape of the in-plane magnetization curve reveals a small in-plane moment, which might be originated from a slight atomic disorder.[28] The out-of-plane magnetization hysteresis loops of the $Mn_{2.7-x}Fe_xGa_{1.3}$ series, measured at 300 K are summarized in Fig. 4 (b). All compositions show clearly that the easy magnetization axis is oriented normal to the film plane.

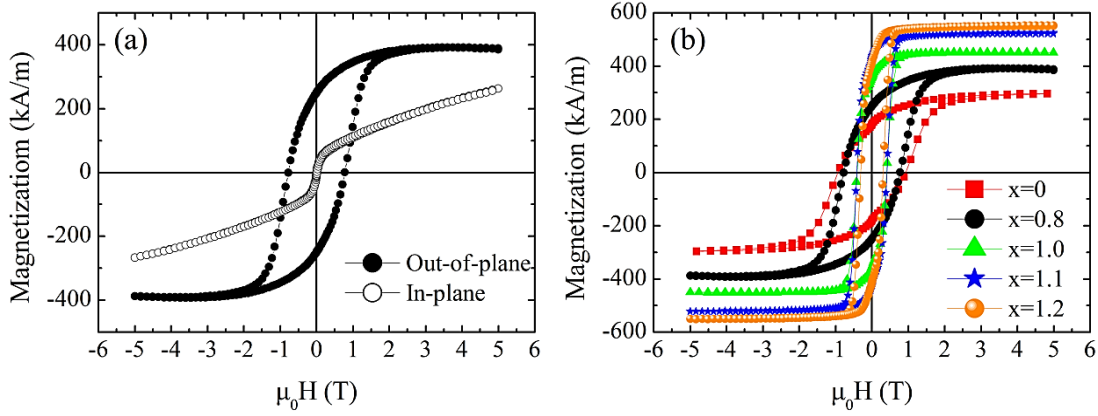

FIG. 4 (a) Out-of-plane and in-plane magnetization hysteresis loops of $Mn_{1.9}Fe_{0.8}Ga_{1.3}$ films. (b) Typical out-of-plane magnetization curves of $Mn_{2.7-x}Fe_xGa_{1.3}$ films, measured at 300 K.

The magnetic properties of $Mn_{2.7-x}Fe_xGa_{1.3}$ films are summarized in Table II. The Fe content x affects the magnetic properties of the tetragonal Mn-Fe-Ga films. The coercive field is reduced from 0.95 T to 0.3 T, while the saturation magnetization is increased from 290 kA/m to 570 kA/m, when the Fe content x increased from 0 to 1.2. The varied magnetization $M_s$ might be related to the chemical disorder and the preferential occupation of the Fe atoms in the $D0_{22}$ unit cell. Fe preferentially substitutes the Mn atoms at the Mn-II site[39] rather than at the Mn-I site, leading to an enhanced $M_s$, due to the reduced antiferromagnetic competition[27,31] between the Mn atoms on the Mn-I and the Mn-II sites. The $Mn_{2.7-x}Fe_xGa_{1.3}$ films show enhanced saturation magnetization, compared to the previously reported Mn-Fe-Ga films with similar compositions (350 kA/m).[35] Furthermore, the Fe content improves the value of the squareness ratio ($S = M_r/M_s$) of out-of-plane loops, from 0.6 for x = 0 to 0.8 for x = 1.1. A further increase of Fe content (x = 1.2) is found to reduce $M_r/M_s$. The effective uniaxial anisotropy $K_u$ is calculated by the equation $K_u = \frac{1}{2}\mu_o M_s H_k$, where $H_k$ is the anisotropy field. The estimated anisotropy field $H_k$ is calculated from linear extrapolation of the in-plane magnetization to the $M_s$ value in the perpendicular direction, since the applied field of 5 T was not sufficient to saturate in-plane the samples. The effective uniaxial anisotropy $K_u$ is larger for the samples containing Fe compared with the $Mn_{2.7}Ga_{1.3}$ film. The highest value of 1.62 kJ/m$^3$ is achieved for the film with x=0.8. Even though $H_k$ decreases with the increase of Fe, the increase of $M_s$ results in large $K_u$ values. The $Mn_{2.7-x}Fe_xGa_{1.3}$ (x=0.8-1.2) films show $K_u$ values larger, compared to $D0_{22}$-$Mn_{3-x}Ga$,[40,41] N-doped $Mn_{2.5}Ga$,[42] tetragonal $Mn_{2.6}Co_{0.3}Ga_{1.1}$,[33] or to $L1_0$-MnAl films.[22,43]

TABLE II: Summary of magnetic properties. Coercive field $\mu_0 H_c$, saturation magnetization $M_s$, squareness ratio $M_r/M_s$, and uniaxial anisotropy $K_u$, taken from the out-of-plane magnetization hysteresis loops of $Mn_{2.7-x}Fe_xGa_{1.3}$ films.

| $Mn_{2.7-x}Fe_xGa_{1.3}$ | $\mu_0 H_c$ (T) | $M_s$ (kA/m) | $M_r/M_s$ | $K_u$ (MJ/m$^3$) |
|---|---|---|---|---|
| x=0 | 0.95 | 290 | 0.60 | 1.35 |
| x=0.8 | 0.78 | 390 | 0.64 | 1.62 |
| x=1 | 0.41 | 450 | 0.78 | 1.50 |
| x=1.1 | 0.40 | 510 | 0.80 | 1.47 |
| x=1.2 | 0.31 | 570 | 0.75 | 1.40 |

Large coercivity values and the maximum energy product *(BH)$_{max}$* are key parameters for permanent magnet applications. *(BH)$_{max}$* is a figure of merit of permanent magnets and proportional to the maximum stored magnetic energy. The out-of-plane *BH*(H) curve of the $Mn_{1.6}Fe_{1.1}Ga_{1.3}$ film and the variation of *(BH)$_{max}$* with the increase of Fe content x are illustrated in Fig. 5 (a) and 5 (b), respectively. The Fe substitution leads to a significant enhancement of the energy product. *(BH)$_{max}$* increases progressively from 8.5 kJ/m$^3$ at x = 0 to 37 kJ/m$^3$ at x=1.1 its value becomes more than quadrupled. The further increase of Fe to x=1.2 reduces *(BH)$_{max}$*, since the coercivity is relatively small. The comparison of the experimentally observed value of the energy product with other reported rare-earth-free permanent magnets, shows that the *(BH)$_{max}$* with the optimized Fe content is larger than that of other tetragonal Mn-based alloys, such as $Mn_xGa$ films (21 or 27 kJ/m$^3$),[21,44] MnAl films (35 kJ/m$^3$),[22] and melt-spun HfCo$_7$ (34 kJ/m$^3$),[17] but smaller than that of the hexagonal MnBi films (130 kJ/m$^3$),[45] LTP MnBi (61 kJ/m$^3$),[46] and $Zr_2Co_{11}$ thin film assemblies (132 kJ/m$^3$).[16] Here, the moderate squareness ratio (*S* = 0.8) limits the magnitude of *(BH)$_{max}$*, which can be further optimized by an appropriate buffer layer.[35]

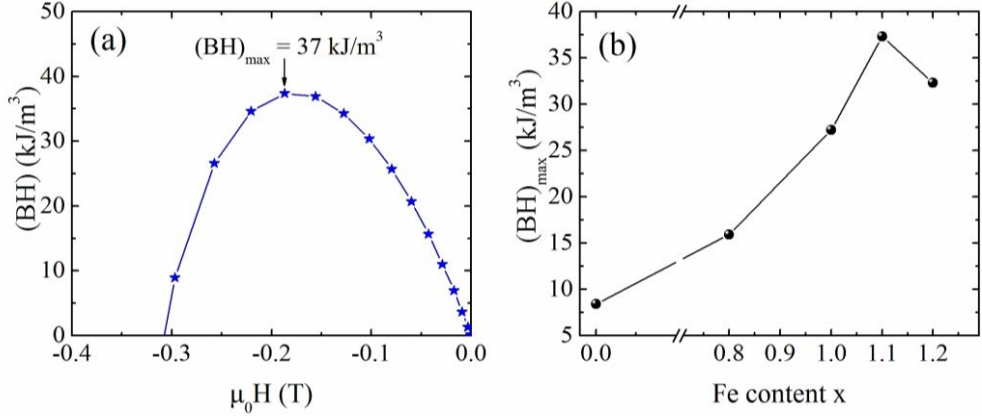

FIG. 5 (a) *BH(H)* curve of the $Mn_{1.6}Fe_{1.1}Ga_{1.3}$ film, and (b) *(BH)$_{max}$* values of $Mn_{2.7-x}Fe_xGa_{1.3}$ films with different Fe content x.

## 4. Conclusions

In summary, we have studied the structural and magnetic properties of $Mn_{2.7-x}Fe_xGa_{1.3}$ films (x = 0-1.2) heteroepitaxially grown on MgO substrates. The films were crystallized in the tetragonal $D0_{22}$ structure with (001) preferred orientation. HRTEM, SAED pattern, and STEM-EDXS analysis revealed that the film is continuous, chemically homogenized, and confirmed furthermore the heteroepitaxial growth. $Mn_{2.7-x}Fe_xGa_{1.3}$ films exhibit strong perpendicular magnetic anisotropy, and tunable magnetic properties depending on the Fe content, with $M_s$ varying from 290 to 570 kA/m, and coercivity from 0.95 down to 0.3 T. This tunable magnetic behavior constitutes an attractive option for spintronic applications. The combination of large $K_u \geq 1.4$ MJ/m$^3$ and high *(BH)$_{max}$* up to 37 kJ/m$^3$ makes $Mn_{2.7-x}Fe_xGa_{1.3}$ films candidate materials for mid-range permanent magnets.

**Acknowledgements**

This work has been funded by the Joint Initiative for Research and Innovation within the Fraunhofer and Max Planck cooperation program. We thank G. Auffermann for the ICP-OES analysis, R. Koban, and W. Schnelle for the magnetic measurements.